\magnification=\magstep1
\tolerance=500
\bigskip
\rightline{1 February, 2023}
\centerline{{\bf ${N\geq 2}$ Particles in the Framework of TeVeS} \footnote{*}{This paper is dedicated to the memory of J.D. Bekenstein (Z''L).}}
\bigskip
\centerline{ L.P. Horwitz}
\bigskip
\centerline{ School of Physics, Tel Aviv University, Ramat Aviv
69978, Israel}
\centerline{ Department of Physics, Bar Ilan University, Ramat Gan
52900, Israel}
\centerline{ Department of Physics, Ariel University, Ariel 40700, Israel}
\bigskip
\noindent{\it Abstract}
\smallskip
\par About 40 years ago, since no viable candidate for ``dark matter'' was discovered, M. Milgrom and J.Bekenstein introduced a non-covariant modification of gravitational theory (MOND) to account for the anomalous rotation curves of galaxies. Bekenstein and Sanders then developed a relativistically covariant of the theory, called TeVeS, involving scaalar and tensor fields. which accounts well for the rotation curves as well for the observed lensing of background radiation around galaxies without the introduction of ``dark matter''. The dynamical behavior of a particle in such a gravitational theory has been recently discussed by Horwitz, Gershon and Schiffer. In this paper we
study the dynamics of the ${N\geq 2}$ particle problem in the framework of the TeVeS theory.
\bigskip
email:larry@tauex.tau.ac.il
\bigskip
\noindent{\bf Introduction}
\smallskip
\par It has been found that galactic rotation curves are not consistent with Newtonian dynamics as they should be for the large interstellar distances in galaxies[1]. To account for this discrepancy, the idea of ``dark matter'' was introduced[2], which permits a reasonable fit to the data.However, no viable candidate for this matter has been found. Milgrom [3] consequently introduced the idea of modifying the gravitational force, with a construction called MOND, which was phenemenologically quite successful, even with relatively universal modification functions, without the need for ``dark matter''.
\par There was, however, no relativistically covariant formulation of the MOND theory until Bekenstein [4] and Bekenstein and Sanders[5] introduced a theory with scalar and tensor fields called TeVeS, successfully accounting for the galactic rotation curves and for the gravitational lensing of light by galaxies from background stars.  
\par Horwitz, Gershon and Schiffer [6] then showed that the TeVeS theory can have a simple Hamiltonian formulation consistent with Einstein relativity through a conformal transformation, for which the potential in the Hamiltonian plays the role of the scalar field of TeVeS.
\par In this paper we discuss the dynamics of $N\geq 2$ particles in this framework. We argue that each particle moves in and independent MOND type gravitational field, but the particles interact in the presence of a Bekenstein-Sanders gauge field [6] which must be non-Abelian. We then show that the resulting geometry can be represented in a Kaluza-Klein structure [7].
\bigskip
\noindent{\bf II. Hamiltonian for ${N\geq 2}$ }
\bigskip
 In Stueckelberg's approach to relativistic dynamics [8], a Hamiltonian of the form ($p^\mu = \eta^{\mu \nu} p_\nu$, $M$ a dimensional parameter)
$$ K ={p_\mu p^\mu \over 2M} + V(x) \eqno(2.1)$$
is defined, for a particle represented as an event, on the manifold ${x^\mu, p_\nu}$, with $V(x)$ a scalar function, invariant under the Lorentz transformations of special relativity. The (Hamilton) equations of motion are then given by\footnote{*}{The dynamics of this theory is discussed extensively in [9]}
$$ {\dot p}_ \mu = - {\partial K \over \partial x^\mu} \eqno(2.2) $$
and
$$ {\dot x^\mu} = {\partial K \over p_\mu}. \eqno(2.3)$$
\par It has recently been shown that this theory can be imbedded in the space-time manifold of general relativity [10].
\par The many body problem [11] can be treated by taking a Hamiltonian of the form
$$ K= \Sigma_{i= 1}^N \{{p_{i\mu}g^{\mu\nu} p_{i\nu} \over 2M_i} + V(x_i)\}, \eqno(2.4)$$
where we assume that each particle is affected independently by the potential energy field.
\par Let us first consider the one-particle case. for which
$$K = {1\over  2M}  p_\mu g^{\mu\nu}(x) p_\nu +V(x), \eqno(2.5)$$
where $V(x)$ is a world scalar field. From the canonical equations,
$$ {\dot p}_\mu = - {1\over 2M}{\partial g^{\lambda \sigma}(x)\over \partial x^\mu} p_\lambda p_\sigma -{\partial V \over \partial x^\mu} \eqno(2.6)$$
and
$${\dot x}^\mu = {1\over M} g^{\mu\sigma}(x)p_\sigma . \eqno(2.7)$$
\par We then have, with the  help of the identity
$$ g_{\beta\sigma} {\partial g^{\beta \gamma} \over \partial x^\mu} g_{\gamma\omega}= - {\partial g_{\sigma \omega} \over \partial x^\mu} ,\eqno(2.8)$$
and its inverse,
the geodesic equation, in the presence of a scalar field,
$$ {\ddot x}= -{\Gamma^\sigma}_{\lambda\nu} {\dot x}^\lambda {\dot x}^\nu - {1\over M}{\partial V \over \partial x^\nu}g^{\sigma \nu}, \eqno(2.9)$$
where
$$ \Gamma^\mu_{\lambda \sigma} = {1 \over 2} g^{\mu\nu}
\bigl\{ {\partial  g_{\nu \sigma} \over \partial x^\lambda} + 
{\partial  g_{\nu \lambda} \over \partial x^\sigma}-{\partial 
  g_{\lambda \sigma} \over \partial x^\nu}\bigr\}$$
is the usual affine connection form [12]
\par For the $N\geq 2$ problem, we assume the additive form
$$ K = K_1 +  K_2 + \dots K_N, \eqno(2.10)$$
where
$$ K_i  =  {1\over  2M_i}  p_{i\mu}g^{\mu\nu}(x_i) p_{i\nu} +V(x_i), \eqno(2.11)$$
We now make a conformal transformation on each of the $K_i$, eliminating the scalar function, by defining
$$ {\hat g^{\mu\nu}(y_i)} = {K_i \over K_i-V(x_i)}g^{\mu\nu} (x_i) \equiv e^{-2\phi(x_i)} g^{\mu\nu}(x_i) \eqno(2.12)$$
\par  Writing this as
$$ {\hat g^{\mu\nu}(y_i)} =  F(y_i)g^{\mu\nu} (x_i) \eqno(2.13)$$
and setting
$$ \delta{ y_i}^\mu= F(y_i) \delta {x_i}^\mu \eqno(2.14)$$
enables us to solve for $y_i(x)$ and the inverse [13]; however
the functions $y_i(x)$ do not cover the whole space time; they are defined only in a patch in the neighborhood of each  of the particles. This is sufficient to define, as for the one particle case, local Einstein geodesics on the submanifold patches $\{y_i\}$. We shall write, with this understanding,
${\hat g^{\mu\nu}(y_i)}\equiv {\tilde g^{\mu\nu}(x_i)}$ in the following.
\par Therefore, a many body Hamiltonian can, in the presence of a world scalar field, acquire a Milgrom type modification of Einstein gravity.
\bigskip
\noindent{\bf III. Many Body Dynamics in TeVeS and Kaluza Klein Structure}
\bigskip
\par We have seen so far that the theory for many particles can treat the particles as essentially independent.  However, we shall see that TeVeS can introduce a non-Abelian gauge field (like the Yang-Mills field [14]) which couples the dynamics of the many body system. In order to achieve a generalized TeVeS theory to account for lensing and gravitational redshifts, Bekenstein and Sanders [5] introduced a timelike vector field ${\cal U}_\mu$  satisfying 
${\cal U}_\mu {\cal U}^\mu = -1$, with ${\cal U}^\mu = g^{\mu\nu}{\cal U}_\nu$. In a
previous work [6] we have considered this field as a gauge field, necessarily non-Abelian to preserve the normalization condition under gauge transformations. With an additional scalar field $\Phi(x,\tau)$, we write the corresponding Hamiltonian for the many body theory  
$$ K=  \Sigma_i^N K_i =\Sigma_i^N \{ {1 \over 2M_i}(p_{i\mu} -\epsilon {\cal U}_\mu (x_i,\tau))g^{\mu\nu}(x_i,\tau)({p_i}_\nu -\epsilon {\cal U}_\nu(x_i,\tau)) + \Phi(x_i,\tau)\} , \eqno(3.1)$$
defined on an $N$ body Hilbert space with direct product basis. Our analysis here is independent of whether the functions are symmetrized (Bose-Einstein statistics) or antisymmetrized (Fermi-Dirac statistics). The scalar field arises as a gauge field to compensate for the $\tau$-derivative of a scalar gauge phase, depending linearly on $\tau$, of each of the wave functions in the Stueckelberg-Schr\"odinger equation, in addition to the ${\cal U}$ fields carrying compensation for the the derivative action of $p_{i\mu}$ on the non-Abelian local phase transformations of the wave functions, {\it i.e.}, under gauge transformations,  each factor $\psi_i (x_i)$ obtains a coefficient $U(x_i,\tau)= U_0(x_i) e^{-i \varphi(x_i)\tau}$, with $U_0(x_i,\tau)$ inducing the non-Abelian gauge. It is easly seen that the $\Phi(x_i,\tau)$ additively can compensate for the Abelian part of the gauge transformation. The space-time derivative ($p_\mu$) introduces, necessarily, a $\tau$ dependence into the gauge fields \footnote{*}{As for the one particle case, one can eliminate the scalar potential terms by defining a modified metric. The Hamilton equations, however, assure that the Hamiltonian remains independent of $\tau$.}:
\par For the non-Abelian part (we suppress the zero subscript),
$$  {\cal U}'(x_i,\tau) = U(x_i)  {\cal U}_\nu(x_i,\tau) U^{-1}(x_i,\tau) - {i\over \epsilon}{\partial U(x_i,\tau) \over \partial {x_i}^\mu} U^{-1}(x_i,\tau) \eqno(3.2)$$
\par As for the Yang-Mills fields, we define the gauge and locally Lorentz covariant force field at each $x$,
$$ f_{\mu\nu} (x,\tau) = {\partial  {\cal U}_\mu \over \partial x^\nu} -
{\partial  {\cal U}_\nu \over \partial x^\mu} + i\epsilon [{\cal U}_\mu,{\cal U}_\nu]. \eqno(3.3)$$
\par The motion of each particle gives rise to a current which, by means of the Yang-Mills propagator [14], generalized to the manifold (a procedure somewhat simplified for weak gravitational fields), contributes to this field in linear superposition. This structure will be studied in a succeeding publication.
\par  We now define a Kaluza-Klein Hamiltonian (with ${\tilde g}^{\mu\nu}(x_i,\tau)$ of the same form as in $(2.12)$ with $V(x_i)$ replaced by $\Phi(x_i.\tau)$) 
$$ K_{KK}= \Sigma_i {\tilde g}^{\mu\nu}(x_i,\tau) p_{i\mu}p_{i\nu}   \eqno(3.4)$$
with the  ($\tau$-dependent) Kaluza-Klein [6][15] metric, at each $x_i$
$$ {\tilde g}^{\mu\nu}(x_i,\tau) = e^{-2\phi(x_i)}(g^{\mu\nu} (x_i,\tau) +  {\cal U}^\mu(x_i,\tau) {\cal U}^\nu(x_i,\tau)) -e^{2\phi(x_i)} {\cal U}^\mu(x_i,\tau) {\cal U}^\nu(x_i,\tau) \eqno(3.5)$$
for which the Hamiltonian has the form
$$K_{KK} = \Sigma_i e^{-2\phi (x_i)}g^{\mu\nu}(x_i, \tau)p_\mu p_\nu - 2\sinh{2\phi (x_i)}({\cal U}^\mu (x_i,\tau)p_\mu)^2 \eqno(3.6) $$
With the definition, on $5D$,  
$$ g^{AB}(x_i,\tau) = \left(\matrix{{\tilde g}^{\mu \nu}(x_i,\tau)& {\cal U}^\nu (x_i,\tau)\cr
{\cal U}^\mu (x_i,\tau)& {\tilde g}^{55}(x_i,\tau)\cr}\right),\eqno(3.7)$$
we have
$$ g^{AB}(x_i,\tau)p_A p_B = {\tilde g}^{\mu \nu}(x_i.\tau) p_{i\mu} p_{i\nu} +2 p_{i5}(p_{i\mu}
{\cal U}^\mu (x_i,\tau)) +{p_{i5}}^2 g^{55}(x_i,\tau). \eqno(3.8)$$
\par Taking
$$ p_{i5}= - {(p_{i\mu}{\cal U}^\mu (x_i,\tau))\over g^{55}(x_i,\tau)}(1 \pm \sqrt{1-2g^{55}(x_i,\tau).\sinh{2\Phi(x_i,\tau)}} , \eqno(3.9)$$
the Kaluza-Klein theory provides an equvalent dynamics generated by
$$ K_{KK}=\Sigma_i {1 \over 2M_i} g_{AB}(x_i, \tau)p_{iA} p_{iB}. \eqno(3.10)$$
\bigskip
\noindent{\bf Conclusions.}
\bigskip
\par In this study, we have shown that a many body dynamics can be formulated in the framework of the $TeVeS$ theory of Bekenstein and Sanders. In this formulation, the Bekenstein-Sanders tensor fields as well as the scalar field arise from the requirement of gauge invariance of the Stueckelberg-Schr\"odinger equation embedded in the manifold of gravitation.   

\bigskip
\centerline {\bf References}
\bigskip
\frenchspacing
[\obeylines
[1] A.Bosma, The Astrophysical Journal {\bf 86} 1825 (1981).
[2] Jan Oort, Bull. Astro. Inst. Neth. {\bf 1},133 (1922);
Fritz Zwicky, Helvetica Physica Acta {\bf 6}, 110 (1933)
[3] M. Milgrom, Astrophysical Jour. {\bf 270}, 384 (1983).
[4] J.D. Bekenstein and M. Milgrom, Astrophys. Jour.{\bf 286}, (1984).
[5] J.D. Bekenstein, Phy. Rev. D {\bf 70},083509 (2004); R.H. Sanders, Astrophys. Jour.{\bf 480}, 492 (1997).
[6] L.P. Horwitz, A. Gershon and M.Schiffer, Found. of Physics {\bf 41}, 14(2011).
[7] A. Gershon and L.P. Horwitz, Jour. Math. Phys.{\bf 50},102704 (2009).
[8] E.C.G. Stueckelberg, Helv. Phys. Acta,{\bf 14}, 588 (1941). 
[9] Lawrence P. Horwitz, {\it Relativistic Quantum Mechanics}, Fundamental Theories of Physics 180, Springer, Dordrecht (2015).
[10] L.P. Horwitz, European Phys. Jour. Plus {\bf 134}, 313 (2019).
[11] Lawrence P. Horwitz and Rafael Arshansky, {\it Relativistic Many-Body Theory and Statistical Mechanics}, IOP Concise Physics, Morgan \& Claypool, Bristol (2018).
[12] S. Weinberg, {\it Cosmology and Gravitation}, Wiley, New York (1972).
[13] L.P. Horwitz,  A. Yahalom,J. Levitan, and M. Lewkovitz, Frontiers of Physics {\bf 12} 124501 (2017).
[14] C.N. Yang and R.L. Mills, Phys. Rev. {\bf 96}, 191 (1954).
[15] A. Gershon and L.P. Horwitz, Jour. Math. Phys. {\bf 50}, 102704 (2009).

\end